\documentclass[sigconf]{acmart}

\usepackage{booktabs}
\usepackage{float}
\usepackage{multirow}
\usepackage{makecell}
\usepackage{algorithm}
\usepackage{algpseudocode}
\usepackage{stfloats}

\newcommand{\revision}[1]{\textcolor{black}{#1}}
\AtBeginDocument{%
  }

\copyrightyear{2026}
\acmYear{2026}
\setcopyright{cc}
\setcctype{by}
\acmConference[UIST '26]{The 39th Annual ACM Symposium on User Interface Software and Technology}{November 02--05, 2026}{Detroit, MI, USA}
\acmBooktitle{The 39th Annual ACM Symposium on User Interface Software and Technology (UIST '26), November 02--05, 2026, Detroit, MI, USA}
\acmDOI{10.1145/3830398.3830518}
\acmISBN{979-8-4007-2856-3/2026/11}
\begin{document}

\title{HeadRoom: Lightweight, Edge-deployable Pipeline for Adaptive Notification Routing}

\author{Dinithi Dissanayake}

\orcid{0009-0007-0178-984X}
\affiliation{%
  \institution{Augmented Human Lab, School of Computing \\ National University of Singapore}
  \city{Singapore}
  \country{Singapore}}
\email{dinithi@ahlab.org}

\author{Prasanth Sasikumar}
\affiliation{%
  \institution{Augmented Human Lab, School of Computing \\ National University of Singapore}
  \city{Singapore}
  \country{Singapore}}
\email{prasanth@ahlab.org}

\author{Suranga Nanayakkara}
\affiliation{%
  \institution{Augmented Human Lab, School of Computing \\ National University of Singapore}
  \city{Singapore}
  \country{Singapore}}
\email{suranga@ahlab.org}

\renewcommand{\shortauthors}{Dissanayake et al.}

\begin{abstract}

Emerging wearables, such as smart glasses, can deliver notifications through multiple sensory channels, but there is still a limited understanding of how to choose the right channel at the right moment. We propose \textsc{HeadRoom}, a lightweight, edge-deployable pipeline that estimates the availability of visual and auditory channels in real time from egocentric video and audio. 
Our controlled user study ($N=25$) shows that, under high perceptual load, routing notifications to the more available channel reduces response time relative to routing them to the less available channel.
%
This work opens up a new possibility for adaptive routing of notifications in wearable and immersive systems.
%

\end{abstract}

\begin{teaserfigure}
\centering
  \includegraphics[width=0.98\textwidth]{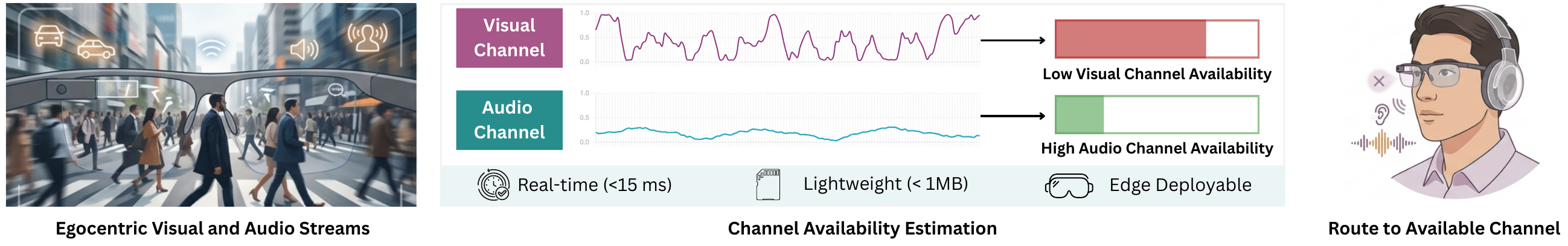}
  \caption{HeadRoom is a lightweight, low-latency pipeline that analyzes egocentric visual and auditory streams in real time to estimate channel availability and route notifications to the more available sensory channel, enabling deployment on wearable devices.}
  \Description{HeadRoom overview. Egocentric visual and auditory streams are analyzed in real time to estimate channel availability using prediction error. The system compares availability across modalities and routes notifications to the more available sensory channel, enabling low-latency, lightweight deployment on wearable devices.}
  \label{fig:teaser}
\end{teaserfigure}


\maketitle

\section{Introduction}
Sensory channels, such as the visual, auditory, and tactile pathways through which people perceive the world.
For effective communication, choosing a sensory channel that is available under the user’s current perceptual load matters
%
~\cite{wickens1983compatibility, wickens2008multiple, proulx2014multisensory}. 



For example, a navigation prompt arriving during a visually demanding moment, or an audio alert in a noisy environment, does not simply add information to the scene. 
When poorly timed or poorly routed, such interventions can disrupt the attentional balance a person had found workable~\cite{kahneman1973attention,bailey2008understanding, dissanayake2026interrupting}. It is well established that overlooking a user’s perceptual state results in missed signals, higher error rates, and a breakdown in flow~\cite{csikszentmihalyi1990flow, dissanayake2025navigating}. These failures might arise from a mismatch between an interface’s chosen output modality and the user’s momentary capacity in the corresponding sensory channel.


In this paper, we study how a multimodal interactive system's chosen output modality disrupts the user. We ground this problem in user’s moment-to-moment capacity: \textit{which sensory channel can absorb new information without disrupting the attentional balance the person is already maintaining?} We term this the \textbf{channel availability problem}.

To operationalize channel availability in real time, we borrow a principle from predictive coding~\cite{friston2009predictive, rao1999predictive}: a sensory channel that is already heavily engaged will produce higher prediction error, because the incoming signal is harder to anticipate. We run lightweight predictors over egocentric visual and audio streams and treat their prediction error as a proxy for how occupied each channel is at any given moment. Higher error means the channel is under load; lower error means it is more available.


We then evaluate our approach in a controlled psychophysical study ($N=25$), where participants view egocentric videos while receiving brief visual and auditory probes under 3 conditions (aligned with our model prediction, random and contradicting with the model prediction).
%
Our findings show that channel availability can be used to adapt routing decisions in a meaningful way and that incorrect routing carries a higher response time. We further demonstrate the model’s feasibility for deployment on off-the-shelf edge hardware.

In summary, our contributions are two-fold:
(1) An open source, lightweight pipeline that estimates channel availability from egocentric audio and video, 
that can be deployed on standard wearable and mobile hardware; and
(2) Controlled empirical evidence demonstrating both the benefits and the boundary conditions of channel-aware routing, together with validation of its feasibility for deployment on resource-constrained wearable devices. Our companion website and pipeline models are available as open-source resources at \url{https://dinithipurna.github.io/HeadRoom/}.

\section{Related Work}
\begin{figure*}[!b]
    \centering
    \includegraphics[width=0.95\linewidth]{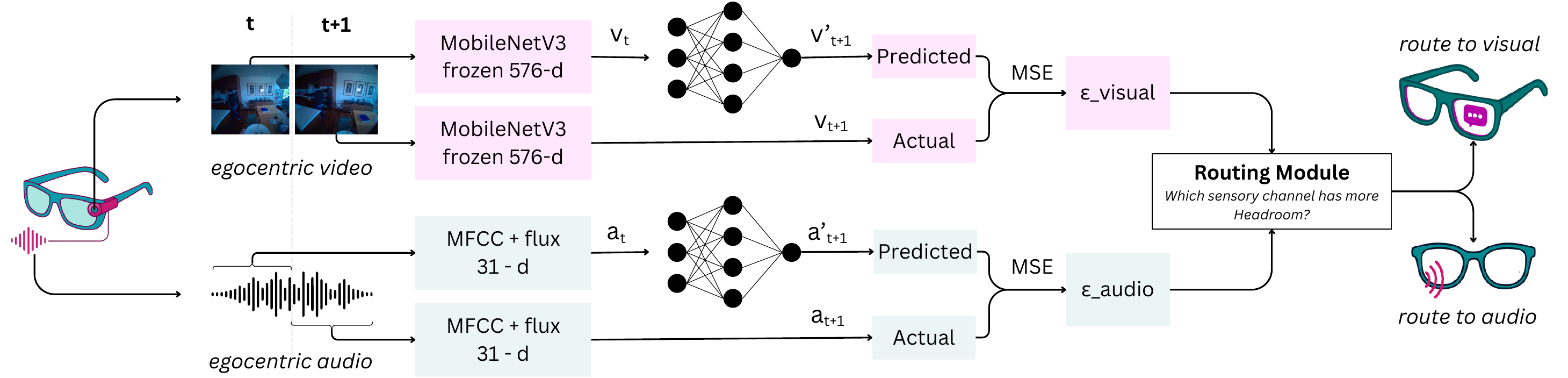}
    \caption{HeadRoom system pipeline. Egocentric video and audio are processed by lightweight predictors, whose errors estimate channel availability for real-time notification routing.}
    \Description{A pipeline diagram showing egocentric video and audio inputs processed by separate visual and audio predictors. Each predictor estimates the next-step features and computes prediction error. The errors are converted into channel availability signals, which are compared by a routing module to decide whether to send notifications to the visual or auditory channel.}
    \label{fig:pipeline}
    \vspace{-20pt}
\end{figure*}

\subsubsection*{\textbf{Sensory Channel Load and Perceptual Surprise}}

Humans can often manage multiple tasks efficiently when those tasks draw on different pools of attentional resources, such as listening to a podcast (auditory) while driving (visual). Multiple Resource Theory (MRT) argues that attention is supported by partly separate resources across modalities and task types \cite{wickens2002multiple, wickens2008multiple}. A key implication of this is that disruption should be greater when concurrent demands compete within the same sensory channel. For example, a visual alert may be more disruptive during a visually demanding moment than an equivalent auditory alert. Related work on modality appropriateness and dual-task interference likewise shows that modality choice affects how well people process concurrent information \cite{welch1980immediate, treisman1973divided, parkes1990route}. Prior research on cognitive load and interruption timing further suggests that attentional capacity fluctuates over time, making some moments more suitable for interruption than others \cite{kahneman1973attention, bailey2008understanding, zheng2026, pu2025promemassist}. Motivated by this perspective, we treat channel availability as a continuous, moment-to-moment property that can differ across sensory channels.

To estimate this property in real time, we draw on predictive coding from neuroscience and psychology. Predictive coding proposes that perception involves continuously comparing incoming sensory input with internal predictions, with larger prediction errors reflecting greater processing demand \cite{rao1999predictive, clark2013whatever}. Following this idea, we use prediction error as a proxy for channel load: high prediction error suggests that a channel is more occupied, whereas low prediction error suggests greater headroom for new information. Because visual and auditory processing draw on partly distinct perceptual resources (
~\cite{wickens1983compatibility, baddeley2013concept}, we model them with two separate predictors rather than a single shared estimate.

\subsubsection*{\textbf{Measuring Perceptual Availability}}

Existing real-time measures of attentional or cognitive load often rely on physiological signals such as pupil dilation, EEG alpha suppression, or galvanic skin response \cite{74_chan2020prompto, 10494102,9974235,wen2024adaptivevoice}. While these signals can provide rich continuous data, they
are usually interpreted as measures of overall cognitive load rather than channel-specific availability. As a result, they are less suited to our goal of comparing sensory channels to decide where a notification should be routed. Behavioral methods, such as secondary-task reaction time measures, provide an alternative \cite{82_arakawa2024prism}, but they are not designed to estimate moment-to-moment perceptual load, 
nor to capture the possibility that the most appropriate output channel may change over time. Other wearable systems rely on rules without any modeling of the user’s immediate perceptual state when making routing decisions \cite{cho2025evaluating, cai2025aiget}.
Our approach occupies a different point in this design space. It uses egocentric audio and video captured by recent wearable devices and produces a continuous estimate of channel availability. 

\section{System Design}

We present \textsc{Headroom}, a lightweight pipeline that estimates perceptual channel availability from egocentric video and audio, and uses this estimate to route notifications. The system has three parts: a visual predictor operating on egocentric RGB frames, an audio predictor operating on short-time acoustic features, and a routing module that compares their prediction errors to decide whether to route to the visual or auditory channel, as illustrated in Figure~\ref{fig:pipeline}. 

\subsection{Problem Formulation}
\label{sec:formulation}

We define \emph{channel availability} as the residual perceptual capacity of a sensory channel at a given moment. Our core idea is motivated by predictive coding theory~\cite{rao1999predictive, friston2009predictive}: when a sensory stream is harder to predict, it is likely to require more processing and therefore leave less spare capacity for new information. We therefore use prediction error as a proxy for channel load. Higher error indicates that a channel is more occupied, while lower error indicates more availability. After normalizing prediction error within each sequence, we define availability as:
$\alpha_t^m = 1 - \hat{\varepsilon}_t^m$
where $\hat{\varepsilon}_t^m \in [0,1]$ is the normalized prediction error for modality $m$ at time $t$.

\subsection{Predictors and Routing Module}

\subsubsection{Visual Predictive Model} For vision, we avoid predicting raw pixels, which would require a larger model and would mostly capture low-level appearance changes. Instead, we predict compact semantic features. Each video frame is resized to $224 \times 224$, normalized using ImageNet statistics~\cite{deng2009imagenet}, and passed through a frozen MobileNetV3-Small backbone~\cite{howard2019searching} with the classification head removed. This produces a 576-dimensional embedding for each frame.
We then train a lightweight two-layer MLP to predict the next-frame embedding from the current embedding. Only this prediction head is trained; the backbone remains frozen throughout. Training is self-supervised and uses consecutive frame pairs from the Aria Everyday Activities dataset~\cite{lv2024aria}. At inference time, the visual prediction error is computed as the mean squared difference between the predicted and observed next-frame embeddings.

\subsubsection{Auditory Predictive Model}
For audio, we use short overlapping windows from mono 16\,kHz recordings. From each window, we extract a compact 31-dimensional feature vector that summarizes spectral shape, energy dynamics, and onset-related properties. Concretely, the features include MFCC statistics~\cite{davis1980comparison}, RMS energy statistics, spectral flux, and zero-crossing rate. These features are designed to capture both gradual acoustic structure and sudden changes in the sound stream.
A lightweight MLP predicts the next audio feature vector from the current one. As in the visual pathway, training is self-supervised and uses consecutive audio windows from the Aria dataset~\cite{lv2024aria}. At inference time, the auditory prediction error is computed as the mean squared difference between the predicted and observed next-window features.

\subsubsection{Routing Module}

Raw visual and auditory prediction errors are not directly comparable, as their scales vary across modalities and scenes. We therefore normalize each error stream in real time using an exponential moving average of its mean and variance. This yields a normalized error in $[0,1]$, which we invert to obtain an availability score, where higher values indicate greater residual capacity.
To support real-time use, we include a short calibration phase at the start of each sequence. During the first 3 seconds, the system updates the normalization statistics without making routing decisions. This allows the availability estimates to adapt to the current environment.
At each time step, routing is determined by comparing the availability of the two channels. The system selects the channel with higher availability, allowing either when the difference is small, and avoids channels that fall below a minimum availability threshold.

\begin{equation}
    m^* =
    \begin{cases}
        \text{audio}  & \text{if } \alpha_t^\text{vis} < \tau \\
        \text{visual} & \text{if } \alpha_t^\text{aud} < \tau \\
        \text{either} & \text{if }
            |\alpha_t^\text{vis} - \alpha_t^\text{aud}| < \delta \\
        \arg\max_m\, \alpha_t^m & \text{otherwise}
    \end{cases}
    \label{eq:routing}
\end{equation}

\noindent Here $\tau$ is the occupancy threshold and $\delta$ is the tie threshold. In our current implementation, we set $\tau = 0.3$ and $\delta = 0.05$ as practical operating values for stable routing. We discuss further details in the Appendix.
\subsubsection{Training}

Both predictors were trained on the same dataset, which contains egocentric recordings captured with Project Aria glasses across everyday activities such as walking, cooking, office work, and social interaction. We used 138 videos (sampled under 10 fps) for training and held out 10\% of the data for validation. In addition, three complete scenarios (6 videos in total) were excluded from training and reserved for the psychophysical study used to evaluate the routing signal.

\section{System Evaluation}


\subsection{Predictive Signal Quality}

Both audio and visual predictors converged stably on held-out data. The visual predictor's validation loss decreased from 0.0515 to 0.0325 by epoch 40, while the audio predictor's validation loss decreased from 0.0191 to 0.00920, reaching its best value at epoch 37. Training and validation losses remained closely matched near convergence, suggesting useful temporal learning without obvious overfitting.

We examined whether the learned availability signals could be explained by simple low-level features such as motion, luminance, and scene cuts. Across both modalities, these associations were generally weak. The strongest correlation was between visual availability and motion energy ($r=-0.309$), which is reasonable given that visual motion is likely to contribute to visual channel load. Overall, this suggests that the prediction-based signal is not dominated by any single handcrafted cue. Instead, it appears to capture a more composite property of the sensory stream. We report the full correlation values in the Appendix.


We also examined \textsc{HeadRoom} in a live streaming setup using Project Aria glasses, where egocentric RGB and audio were streamed to a laptop for real-time inference. To stress different channels, we created simple scenarios involving abrupt sounds, visually busy or high-motion scenes, and calmer periods in which both channels were relatively stable. The resulting availability traces generally shifted in intuitive directions: abrupt acoustic events reduced auditory availability, visually dynamic scenes reduced visual availability, and quieter moments often produced more balanced signals. Although preliminary, these observations suggest that the learned signal remains responsive in live use and behaves broadly consistently with moment-to-moment changes in sensory demand. We provide one such qualitative example from a live Aria streaming setup on our interactive companion website.



\subsection{Edge Feasibility}

\subsubsection{Runtime and Resource Footprint}


\begin{table}[h]
\centering
\renewcommand{\arraystretch}{0.9}
\caption{Runtime characteristics of the \textsc{HeadRoom} prototype on Meta Quest 3S headset.}
\begin{tabular}{lr}
\toprule
Metric & Value \\
\midrule
Mean latency/step & 10.98\,ms \\
Median latency/step & 10.08\,ms \\
95th percentile & 12.75\,ms \\
Memory overhead & 10\,MB \\
\revision{Prediction-head checkpoint size} & 0.625\,MB \\ 
\revision{Full ONNX deployment size} & \revision{4.4\,MB} \\
\bottomrule
\end{tabular}
\Description{This table summarizes runtime characteristics of the HeadRoom prototype on a Meta Quest 3S headset. The system runs with a mean latency of 10.98 ms per step, a median latency of 10.08 ms, and a 95th percentile latency of 12.75 ms, indicating consistently low inference delay. The prototype uses 10 MB of memory overhead, and the total model checkpoint size is 0.625 MB, suggesting a lightweight design suitable for on-device deployment.}
\label{tab:runtime_headset}
\end{table}

We benchmarked \textsc{HeadRoom} on an Apple M3 Pro to assess runtime feasibility. The full end to end system runs comfortably in real time on CPU, with a mean latency of 43.1\,ms per step and a throughput of 23.2 steps/s, which is more than twice the rate required for our 10\,fps setting. The model footprint is also small, with a total checkpoint size of only 0.625\,MB. GPU (MPS) execution further reduced latency to 12.5\,ms per step and increased throughput to 79.8 steps/s, suggesting that the pipeline is lightweight enough for real-time use across both laptop-class CPU and GPU settings.

\subsubsection{Edge Deployment}

To assess deployment feasibility, we implemented a lightweight standalone prototype of \textsc{HeadRoom} within a Unity application and executed it on a Meta Quest~3S, an XR headset with 8\,GB of onboard memory and a Snapdragon XR2 Gen~2B chipset. We exported the trained PyTorch components to ONNX\footnote{https://docs.unity3d.com/Packages/com.unity.barracuda@1.0/manual/Exporting.html} and integrated them using Unity's local inference engine, removing the need for an external laptop or Python server. This prototype includes the MobileNetV3 backbone together with the visual and audio MLP predictors, and logs timestamped inference outputs directly on the device during runtime.
Across approximately 180 inference steps, the prototype achieved a mean latency of 10.98\,ms per step (median: 10.08\,ms, 95th percentile: 12.75\,ms) as shown in Table~\ref{tab:runtime_headset}. These results suggest that the core \textsc{HeadRoom} pipeline is lightweight enough to support practical real-time routing in a contemporary wearable setting.

\section{User 
Study}
\label{sec:study}

\subsection{Overview and Method}
\label{sec:overview}

We conducted a psychophysical study to examine whether \textsc{HeadRoom}'s routing signal corresponds to measurable perceptual performance. Participants watched three egocentric scenarios and received visual and auditory probes under three routing conditions: \textsc{Model}, \textsc{Inverse}, and \textsc{Random}. In \textsc{Model}, probes were routed to the predicted more available channel; in \textsc{Inverse}, probes were routed to the predicted less available channel; and in \textsc{Random}, probes were routed randomly without using our model. Each participant viewed all three scenarios, with condition-to-scenario assignment counterbalanced across participants using all six permutations. Each scenario contained 40 probes, yielding 120 trials per participant. After each clip, participants answered a comprehension question and completed a raw NASA-TLX rating~\cite{hart1988development}.


We recruited $N=25$ participants (7 female, 18 male; age $M=30.08$, $SD=7.11$). All reported normal or corrected-to-normal vision and hearing. The study was implemented in Unity\footnote{https://unity.com/} and presented on a Meta Quest 3S\footnote{https://www.meta.com/quest/quest-3s/} headset. Visual probes were brief white squares (200\,ms) presented in the upper-right region of the display, while auditory probes were 1\,kHz tones, also 200\,ms in duration, delivered through the headset speakers. These stimuli were designed as simple, high-contrast signals that could be rapidly detected without requiring semantic processing~\cite{macdonald2011visual, neville1987attention}. The short duration allows us to probe moment-to-moment perceptual availability while minimizing interference with the ongoing task~\cite{zhao2022robust}. Participants responded with the right controller index trigger whenever they detected a probe. They completed a short practice block and were required to achieve at least 80\% detection accuracy before continuing to the main task. Probe onsets and response times were logged in Unix milliseconds.


We used three held-out egocentric scenarios from the Aria Everyday Activities dataset to represent a range of perceptual demands. The videos were selected through manual browsing based on their apparent activity level and sensory complexity, and were excluded from model training. Video~1 depicts moderate everyday activity, Video~2 a lower-demand cooking scenario, and Video~3 a higher-demand scene with dynamic motion and concurrent speech. All clips were approximately 6 minutes long and were presented on a large virtual screen in the headset. Participants were instructed to remain attentive throughout each scenario and answer a single comprehension question afterward. 
Our companion website shows the three videos and the example probe placements.

\subsection{Key Findings}
\label{sec:findings}

Three participants were excluded for failing to maintain the 80\% detection threshold required during the training phase, leaving a final sample of $N=22$. Responses were matched to probe onsets within a 100--1500\,ms window. Trials with no valid response in this window were treated as misses, and mean reaction time (RT) was computed from valid responses only. Detection rates were high overall. Auditory probes were detected almost perfectly, whereas visual detection was lower (around 86\%), but detection did not differ significantly across routing conditions. Across all videos and conditions, auditory probes were responded to more slowly than visual probes (audio: $M=636$\,ms, visual: 
$M=574$\,ms; $t=5.632$, $p<.001$, $d=1.201$).

\subsubsection{\textbf{Routing Effects Depended on Scene Demand.}}
\begin{figure}[h]
    \centering
    \includegraphics[width=\linewidth]{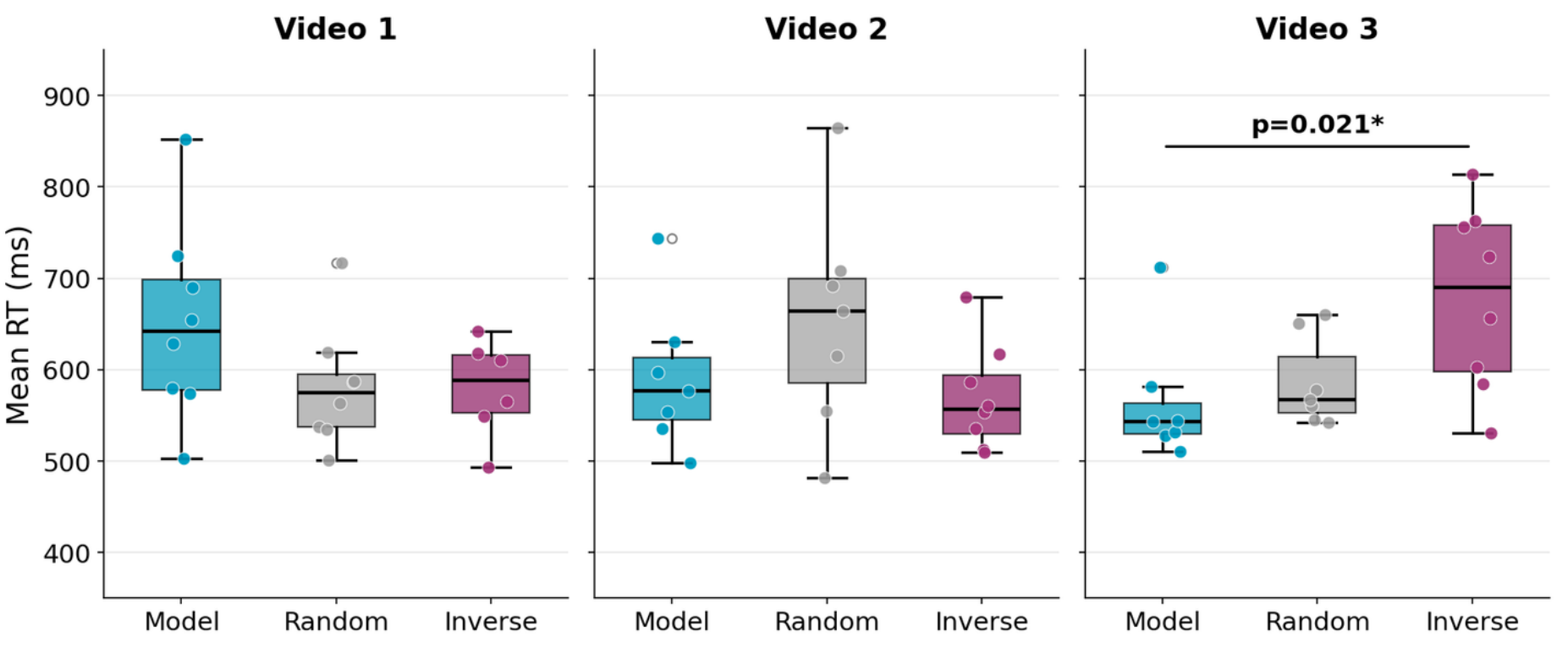}
    \caption{Mean response times for \textsc{Model}, \textsc{Random}, and \textsc{Inverse}. The only significant difference appeared in video 3, where \textsc{Model} is faster than \textsc{Inverse}, suggesting that routing to the more available channel reduces response time under high demand.}
    \Description{Boxplots of mean response time across Model, Random, and Inverse conditions for three videos. Only Video 3 shows a significant difference, with Model faster than Inverse.}
    \label{fig:graphresults}
\end{figure}

Pooled across all three videos, the routing manipulation produced a small directional trend (Model vs. Inverse mean difference $= -8$\,ms, $d = -0.17$; 16 of 22 participants faster under Model), but this did not reach conventional significance on either the Wilcoxon signed-rank test ($W=83$, $p=.166$) or the paired $t$-test ($p=.435$). This pooled null is consistent with our expectations: Videos~1 and~2 involved lower-moderate perceptual demand and offered the routing signal less opportunity to differentiate channels. The more informative test is Video~3, which was designed to place the highest perceptual demand on participants, as later confirmed by comprehension accuracy dropping to 50\% 
compared with ceiling-level performance on Videos~1 and~2 (Figure~\ref{fig:graphresults}). This is consistent with prior work suggesting that routing decisions matter most when perceptual demands increase competition for limited resources~\cite{wickens2008multiple, pu2025promemassist}. 

\subsubsection{\textbf{\textsc{HeadRoom} Reduced Response Time Under High Demand.}}
Video~3 showed the clearest significant condition-dependent pattern. A Kruskal--Wallis test revealed a significant difference across routing conditions ($H=7.10$, $p=.029$), with responses fastest under Model routing and slowest under Inverse, with Random falling between (mean ranks: Model $=7.14$, Random $=10.71$, Inverse $=16.00$). The Model condition was significantly faster than Inverse (difference $=-114$\,ms, two-tailed $p=.021$, $r=.598$, $d=-1.31$) which is a large effect. Random likewise trended faster than Inverse (difference $=-93$\,ms, $p=.072$), while Model and Random did not differ significantly (difference $=-22$\,ms, $p=.209$). \revision{As descriptive follow-up analyses, Holm-corrected pairwise comparisons showed the same ordering of conditions, although no individual comparison remained statistically significant after correction. We therefore use the trial-level linear mixed-effects model below as the primary inferential analysis.}

Linear mixed-effects models on log RT with random intercepts for participants confirmed this pattern. Using Inverse as the reference condition, both Model ($\beta=-0.178$, $p=.003$) and Random ($\beta=-0.146$, $p=.013$) produced significantly faster responses. Adding channel as a fixed effect did not change this picture: Model tended to be 13\%  faster in RT than Inverse ($\beta=-0.132$, $p=.022$), and Random also remained faster than Inverse ($\beta=-0.114$, $p=.048$). Together, these results indicate that routing notifications to the less available channel measurably increased response cost under high perceptual demand. We provide the complete mixed model results in the Appendix.

\revision{As a robustness check, we examined whether simple scene cues could explain the observed effects. Appendix Table~\ref{tab:feature_correlations} shows that the estimated channel availability was not strongly correlated with any single low-level feature. In addition, neither motion energy ($\beta=0.445$, $p=.291$) nor spectral flux ($\beta=0.007$, $p=.390$) significantly predicted trial-level log RT in Video~3. Together, these results suggest that the response-cost pattern was not explained solely by visual motion or acoustic change.}

NASA-TLX ratings did not differ significantly across routing conditions. This dissociation between response time and subjective workload suggests that channel-aware routing affects fine-grained perceptual processing without altering participants' global sense of effort. This is consistent with the view that availability operates below the threshold of conscious load appraisal.

\subsubsection{\textbf{The Availability Signal Tracked Behaviourally Meaningful Variation.}} To examine whether the availability signal captured genuine moment-to-moment perceptual structure rather than arbitrary model fluctuations, we correlated trial-level availability estimates with RT in Video~3. Higher availability of the probed channel was associated with faster responses (Pearson $r=-0.112$, $p=.002$; Spearman $\rho=-0.136$, $p<.001$; bootstrap 95\% CI $[-0.182, -0.043]$). This relationship was stronger for \emph{availability delta}, which is the difference in availability between the probed and alternative channel at each moment. Availability delta showed a larger negative association with RT (Pearson $r=-0.201$, $p<.001$; Spearman $\rho=-0.210$, $p<.001$; bootstrap 95\% CI $[-0.267, -0.131]$), indicating that participants responded fastest not merely when the probed channel was available in isolation, but when it was \emph{relatively more available} than the competing channel. This is precisely the quantity \textsc{HeadRoom} uses to make routing decisions, and its behavioural relevance supports the core premise of our approach.

\section{Discussion and Limitations}

\paragraph{Implications for Timing and Modality}
Our results suggest that channel availability may be useful not only for deciding \emph{which} modality to use, but also \emph{when} to interrupt. We showed that lower availability was associated with slower responses, and relative channel availability showed an even stronger relationship with RT. This suggests that channel availability may support not only modality selection but also in deciding when to interrupt. In this sense, our findings are broadly consistent with systems such as Promemassist~\cite{pu2025promemassist}, which reduce interruption cost by modeling working memory, while our \textsc{HeadRoom} remains a lightweight alternative. This timing signal may be especially useful when modality choice is limited (e.g., only audio output is available) or fixed (e.g., a notification must be displayed visually), as it can help determine whether the notification should be delayed until a more appropriate moment.

\paragraph{Beyond Audio and Vision}
We focused on visual and auditory output because they are common channels for wearable notifications and allowed us to test the core idea of channel-specific availability. However, many wearable systems also support haptic feedback\cite{ParametricHaptics, 2026llmglasses}, which could provide an alternative when both visual and auditory channels are occupied. Extending the framework to include haptics is, therefore, a natural next step. 
More broadly, not all information is equally well suited to every modality, and this remains a limitation of the current formulation. 

\paragraph{Potential Use Cases}
The observed benefits were on the order of milliseconds for brief probes, but such differences may still matter in settings where timely responses are important. Attention-aware routing could, for example, support first responders, drivers, or other users operating in demanding environments by reducing the disruption caused by poorly timed or poorly routed alerts.  It may also enable new interaction designs in games and immersive applications, where routing to a less available channel could increase challenge, while routing to a more available channel could reduce disruption. However, these benefits should be interpreted in light of our task: we measured rapid detection of brief probes, and the effects may differ for richer notifications that require deeper processing, interpretation, or action.

\paragraph{Limitations and Future Work}
Our approach has several limitations. First, prediction error is only a proxy for perceptual availability. While we chose it for its lightweight, self-supervised nature and suitability for real-time use, more expressive or structured predictors may yield stronger estimates.  

Second, our evaluation used a highly controlled probe-detection task. Participants responded to brief probes while passively viewing egocentric videos, with visual probes fixed in location and timing. This design helped us isolate routing effects, but it does not fully capture how notifications are experienced in everyday wearable use~\cite{zheng2026}, and it constrained our evaluation to rapid detection rather than deeper cognitive processing. One consequence of this setup was that responses to visual probes were systematically faster than responses to auditory probes, likely because participants were already visually engaged with the screen and thus better primed for visual events, rather than because visual delivery was genuinely more effective. As a result, although our findings suggest that routing to the less available channel can increase response time, they do not yet establish a clear advantage of model-based routing over a simpler random policy in this controlled setting. Future work should also evaluate the approach in more ecologically valid tasks that require interpretation, memory, or action.

\revision{In addition, our evaluation used three manually selected scenarios rather than a formally validated stimulus set. Although selected a priori to span different levels of sensory demand, subsequent behavioral analyses supported this ordering: Video 3 had 50\% comprehension accuracy, compared with ceiling-level performance for Videos 1 and 2, and significantly higher response-time variance (Levene's test: $W = 3.54$, $p = .029$). Future work should use systematically validated stimuli spanning a broader range of real-world activities and sensory demands.}

Third, although our framework models visual and auditory availability separately, these channels are not fully independent in real use. Future work could explore models that capture both channel-specific effects and cross-modal dependencies, and examine how this interplay informs both timing and modality selection. \revision{We also evaluated HeadRoom primarily against random and inverse routing to establish the feasibility of perceptual availability as a routing signal. Future work should benchmark additional routing strategies, including fixed-modality and heuristic policies. We have released trial-level reaction times, routing logs, and availability traces with our code repo to facilitate such comparisons.} Finally, perceptual availability is likely influenced not only by external sensory input but also by internal factors such as fatigue, expectation, task goals, and stress, which were not modeled here.

\section{Conclusion}

We introduced \textsc{HeadRoom}, a lightweight and edge-oriented approach for estimating moment-to-moment visual and auditory channel availability from egocentric sensing. Our controlled psychophysical study, suggested that channel availability provides a useful basis for adaptive notification routing: directing notifications to the relatively less available channel can increase response time.
These findings highlight channel availability as a promising design primitive for future wearable and immersive systems that aim to deliver information with less disruption and greater contextual sensitivity.

\begin{acks}


We thank Ryo Hajika for his valuable feedback on the study design and use case. We also thank Chitralekha Gupta, Hyung Woon Lee, and Yize Wei for their helpful comments on earlier drafts of this paper, and our lab members for their assistance with pilot testing the interfaces.


\end{acks}

\balance
\bibliographystyle{ACM-Reference-Format}
\bibliography{references}

\appendix
\section{Appendix}

We have open-sourced all our models, including the Unity-compatible ONNX versions and trial-level RT data. Our companion website and pipeline resources are available at \url{https://dinithipurna.github.io/HeadRoom/}. The website includes examples of the probes used in our study, snippets from the study videos, and an interactive qualitative demo illustrating the model’s real-time behavior. This appendix further details the normalization component of our pipeline, the study procedure, and the complete study results.

\subsection{Modeling Channel Availability}

For each channel, we convert the raw prediction error into an availability score using an online per-sequence normalizer as shown in Algorithm~\ref{alg:availability_norm}. During an initial warm-up period of 30 steps, we collect raw errors and return a neutral score of 0.5. Once calibrated, we maintain an exponentially weighted running mean and variance of the error stream, compute a z-score, clip it into a bounded range, and invert it so that higher prediction error corresponds to lower channel availability. For audio, we first apply a $\log(1+1000e_t)$ compression to reduce the effect of rare extreme spikes. Finally, we floor the availability at 0.05 to avoid unstable hard-zero behavior. 

The parameters $\tau=0.3$ and $\delta=0.05$ were set a priori, but their influence on the present study is limited. Because probes were only selected at moments where availability was at least $0.3$, the threshold mainly acted as a minimum filter rather than a strong experimental manipulation. In contrast, post hoc exploration suggested that routing benefits weakened when the threshold was reduced to $0.2$ or below, indicating that stricter validation of the threshold choice may be needed in future work. By comparison, $\delta$ may have had little practical effect in this study, since the \textit{either} cases often still showed a slight visual advantage. Future work should therefore more systematically examine the sensitivity of the results to these parameters.

\begin{algorithm}[H]
\caption{Normalization of channel availability}
\label{alg:availability_norm}
\footnotesize
\begin{algorithmic}[1]
\Require Raw prediction error $e_t$ for a channel at time $t$
\Require Running mean rate $\alpha_\mu$, running variance rate $\alpha_{\sigma}$, clipping range $c$
\Require Warm-up length $T_{\mathrm{warm}}=30$, minimum availability floor $\epsilon=0.05$
\Statex
\State \textbf{State:} running mean $\mu_t$, running variance $\sigma_t^2$, calibration buffer $\mathcal{B}$

\If{channel is audio}
    \State $e_t \leftarrow \log(1 + 1000\, e_t)$
\EndIf

\If{normalizer is not yet calibrated}
    \State Append $e_t$ to $\mathcal{B}$
    \If{$|\mathcal{B}| \geq T_{\mathrm{warm}}$}
        \State Initialize
        \[
        \mu_t \leftarrow \mathrm{mean}(\mathcal{B}), \qquad
        \sigma_t^2 \leftarrow \mathrm{var}(\mathcal{B}) + 10^{-6}
        \]
        \State Mark normalizer as calibrated
    \EndIf
    \State \Return $0.5$
\EndIf

\State $\mu_t \leftarrow \alpha_\mu e_t + (1-\alpha_\mu)\mu_{t-1}$
\State $r_t \leftarrow (e_t - \mu_t)^2$
\State $\sigma_t^2 \leftarrow \alpha_{\sigma} r_t + (1-\alpha_{\sigma})\sigma_{t-1}^2$
\State $\sigma_t \leftarrow \max\left(\sqrt{\sigma_t^2}, 10^{-6}\right)$
\State $z_t \leftarrow \dfrac{e_t - \mu_t}{\sigma_t}$
\State $n_t \leftarrow \mathrm{clip}\left(\dfrac{z_t + c}{2c},\, 0,\, 1\right)$
\State $a_t \leftarrow \max(\epsilon,\; 1 - n_t)$
\State \Return $a_t$
\end{algorithmic}
\end{algorithm}

\subsection{User Study Procedure}

The user study was conducted entirely on a VR headset in a quiet, secluded room to support immersion and minimize external distraction. At the start of each session, the participant selected a trial number corresponding to one of the six counterbalanced condition permutations. The study then began with a short calibration phase presented on a black screen, during which participants responded to visual and auditory probes presented with the same visual location and sound settings used in the main task. Participants were required to reach at least 80\% accuracy in this phase before proceeding.

After calibration, participants completed the main study, which consisted of three videos presented in the same order, with routing conditions assigned according to the selected trial permutation. During each video, participants watched the content and responded to the probes using the right-hand controller. Their primary task was to attend to the video well enough to answer a comprehension question afterward, while also responding as quickly as possible to the probes. After each video, participants completed the comprehension question and a raw NASA-TLX questionnaire.

Probe moments were selected by dividing each video into 40 bins and choosing one random probe time per bin, subject to the constraint that the predicted more available channel had an availability of at least 0.3.

\begin{figure}[h]
    \centering
    \includegraphics[width=0.6\linewidth]{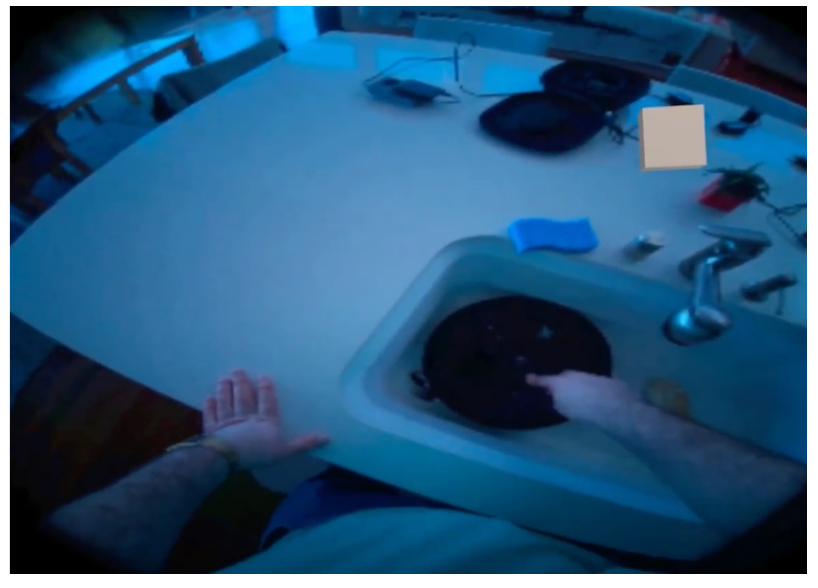}
    \caption{Visual probe - square in the upper-right corner, used in the calibration and main study.}
    \Description{A figure showing the visual probe, which is a square appearing in the upper-right corner, used in the calibration and main study.}
    \label{fig:visual_probe}
\end{figure}

\begin{figure}[h]
    \centering
    \includegraphics[width=0.9\linewidth]{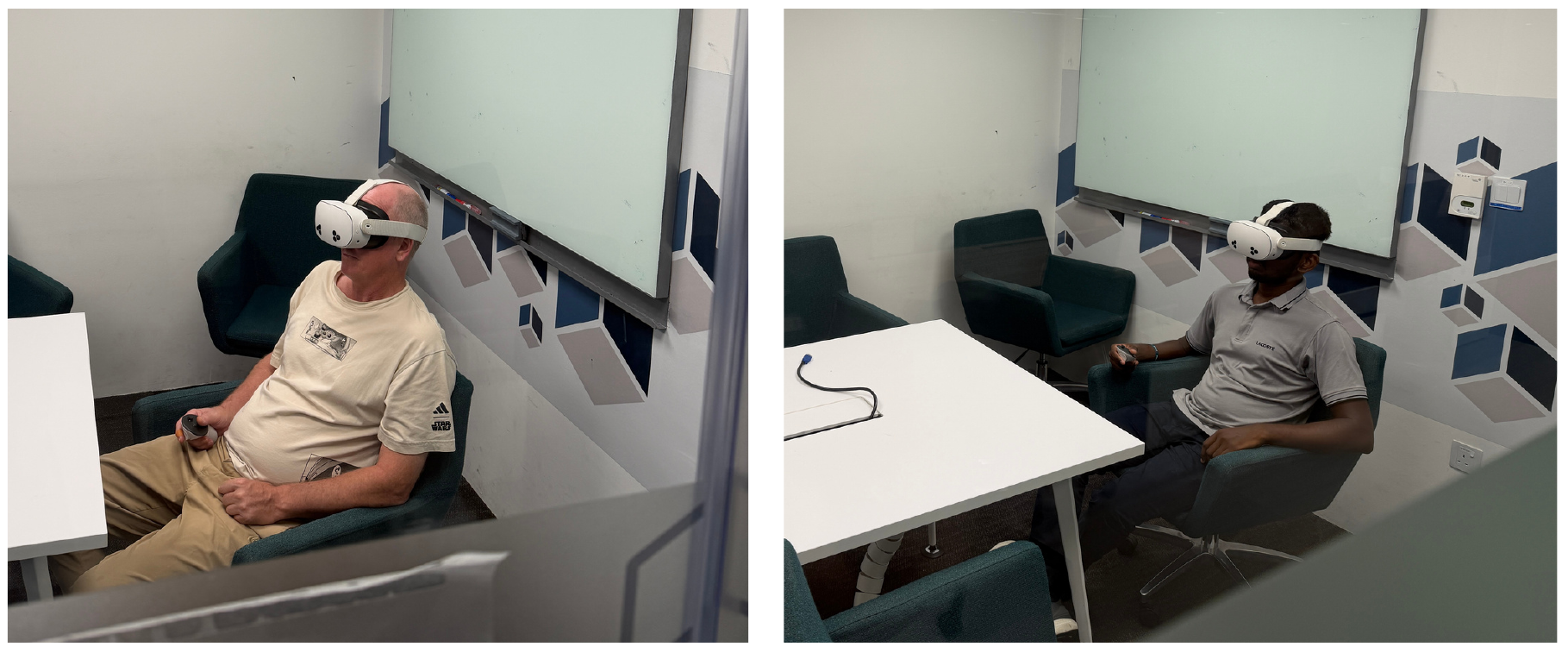}
    \caption{Participants completing the study in the headset and responding to probes using the right-hand controller.}
    \Description{Two participants are in a separate room completing the study in the headset and responding to probes using the right-hand controller.}
    \label{fig:user_study_participants}
\end{figure}

The research team did not interrupt participants during video viewing. Figure~\ref{fig:visual_probe} illustrates the probe design, and Figure~\ref{fig:user_study_participants} shows participants completing the study while viewing the videos and responding to probes in the headset in a separate quiet room.


\revision{\subsection{Channel Availability Statistics}}

\revision{Table~\ref{tab: availabitly_Stats} reports the probe-level channel availability at all scheduled probe moments. Probe times were selected randomly from timestamps satisfying the scheduling criterion (selected channel availability $\geq \tau=0.3$); they were not manually chosen to disadvantage the Inverse condition.}

\begin{table}[h]
\centering
\caption{\revision{Probe-level channel availability (mean $\pm$ SD).}}
\begin{tabular}{lcc}
\toprule
Video & Model (chosen) & Inverse (non-chosen) \\
\midrule
Video 1 & 0.629 $\pm$ 0.149 & 0.410 $\pm$ 0.205 \\
Video 2 & 0.694 $\pm$ 0.140 & 0.450 $\pm$ 0.196 \\
Video 3 & 0.663 $\pm$ 0.178 & 0.439 $\pm$ 0.186 \\
\bottomrule
\end{tabular}
\Description{Table reporting mean and standard deviation of probe-level channel availability for Model-selected and Inverse-selected channels across Videos 1, 2, and 3. Model-selected channel availability ranged from 0.629 to 0.694, while Inverse-selected channel availability ranged from 0.410 to 0.450. Although the Inverse condition used the relatively less available channel, its mean availability remained above the scheduling threshold of 0.3 in all three videos.}
\label{tab: availabitly_Stats}
\end{table}

\revision{Although the Inverse condition received the relatively less available channel, the non-chosen channel still exhibited moderate availability across all videos (mean = 0.41--0.45, above $\tau=0.3$), indicating that probes were not scheduled at moments when the alternative channel was maximally occupied. You can find the raw availability scores at the probed moments at our code repository.}

\subsection{Study Results}
\noindent \textbf{Overall Pooled Results - Within Subjects}

Across all conditions, auditory probes produced significantly slower responses than visual probes ($t=5.63$, $p<.001$, $d=1.20$), with mean response times of 636\,ms and 574\,ms, respectively. The model condition was faster than the inverse condition on 16 out of 22 participants, hence being significant in the sign test~\ref{tab:overall_paired}.

\begin{table}[H]
\centering
\footnotesize
\renewcommand{\arraystretch}{0.95}
\caption{Overall paired comparisons across routing conditions.}
\begin{tabular}{lcccc}
\toprule
\makecell[l]{Comparison} &
\makecell[c]{Mean diff\\(ms)} &
\makecell[c]{Wilcoxon\\($W$, $p$ two-tail)} &
\makecell[c]{Sign\\test} &
\makecell[c]{Paired\\$t$-test ($p$)} \\
\midrule
Model vs.\ Inverse  & -8 & $83$, $.166$  & $16/22$, $.026$ & $.435$ \\
Model vs.\ Random   & -2 & $124$, $.949$ & $11/22$, $.584$ & $.904$ \\
Random vs.\ Inverse & -6 & $94$, $.305$  & $13/22$, $.262$ & $.380$ \\
\bottomrule
\end{tabular}
\Description{Paired within-subject comparisons show very small mean response-time differences across routing conditions. Model is 8 ms faster than Inverse, 2 ms faster than Random, and Random is 6 ms faster than Inverse. Wilcoxon sign tests and paired t-tests show no statistically significant pairwise differences.}
\label{tab:overall_paired}
\end{table}

\noindent\textbf{Per-Video Between-Subjects Response Time Analysis}

Per-video analyses are reported in Tables~\ref{tab:video1_between}, ~\ref{tab:video2_between}, and \ref{tab:video3_between}. Only Video~3, the highest-demand scenario, showed a significant effect of routing condition, with \textsc{Model} producing faster responses than \textsc{Inverse}, while Videos~1 and~2 showed no significant differences.

\begin{table}[H]
\centering
\small
\renewcommand{\arraystretch}{0.95}
\caption{Between-subjects response times for Video 1 (Moderate).}
\begin{tabular}{lcccc}
\toprule
Condition & $n$ & Mean (ms) & SD (ms) & Range (ms) \\
\midrule
Model   & 8 & 651 & 107 & [503, 852] \\
Random  & 8 & 581 & 66  & [501, 717] \\
Inverse & 6 & 580 & 55  & [493, 642] \\
\midrule
\multicolumn{5}{l}{Kruskal--Wallis: $H=3.158$, $p=.206$} \\
\multicolumn{5}{l}{Model vs.\ Inverse: $p=.142$} \\
\multicolumn{5}{l}{Model vs.\ Random: $p=.161$} \\
\multicolumn{5}{l}{Random vs.\ Inverse: $p=.852$} \\
\bottomrule
\end{tabular}
\Description{For Video 1, mean response times are 651 ms for Model, 581 ms for Random, and 580 ms for Inverse, with similar variability across groups. A Kruskal–Wallis test shows no significant difference between routing conditions, and all pairwise comparisons are non-significant.}
\label{tab:video1_between}
\end{table}

\begin{table}[H]
\centering
\small
\renewcommand{\arraystretch}{0.95}
\caption{Between-subjects response times for Video 2 (Easy-Cooking).}
\begin{tabular}{lcccc}
\toprule
Condition & $n$ & Mean (ms) & SD (ms) & Range (ms) \\
\midrule
Model   & 7 & 591 & 80  & [498, 744] \\
Random  & 7 & 654 & 122 & [482, 865] \\
Inverse & 8 & 569 & 57  & [510, 680] \\
\midrule
\multicolumn{5}{l}{Kruskal--Wallis: $H=2.491$, $p=.288$} \\
\multicolumn{5}{l}{Model vs.\ Inverse: $p=.694$} \\
\multicolumn{5}{l}{Model vs.\ Random: $p=.318$} \\
\multicolumn{5}{l}{Random vs.\ Inverse: $p=.152$} \\
\bottomrule
\end{tabular}
\Description{For Video 2, mean response times are 591 ms for Model, 654 ms for Random, and 569 ms for Inverse. The overall Kruskal–Wallis test is not significant, and none of the pairwise comparisons reach significance.}
\label{tab:video2_between}
\end{table}

\begin{table}[H]
\centering
\small
\renewcommand{\arraystretch}{0.95}
\caption{Between-subjects response times for Video 3 (Hard-Movement).}
\begin{tabular}{lcccc}
\toprule
Condition & $n$ & Mean (ms) & SD (ms) & Range (ms) \\
\midrule
Model   & 7 & 564 & 69  & [510, 712] \\
Random  & 7 & 586 & 49  & [542, 660] \\
Inverse & 8 & 679 & 100 & [530, 814] \\
\midrule
\multicolumn{5}{l}{Kruskal--Wallis: $H=7.096$, $p=.029$} \\
\multicolumn{5}{l}{Model vs.\ Inverse: $p=.021$} \\
\multicolumn{5}{l}{Model vs.\ Random: $p=.209$} \\
\multicolumn{5}{l}{Random vs.\ Inverse: $p=.072$} \\
\bottomrule
\end{tabular}
\Description{For Video 3, mean response times are 564 ms for Model, 586 ms for Random, and 679 ms for Inverse. The overall Kruskal–Wallis test is significant, indicating differences across routing conditions. Pairwise testing shows that Model is significantly faster than Inverse, while Model vs. Random and Random vs. Inverse are not significant.}
\label{tab:video3_between}
\end{table}

\newpage

\noindent\textbf{Mixed Model Results for Video 3}

For Video~3, mixed-effects models (Table~\ref{tab:video3_mixedlm}) confirmed a significant policy effect on log-transformed response time. Relative to \textsc{Inverse}, both \textsc{Model} and \textsc{Random} produced faster responses, and this pattern remained significant after adjusting for probe channel. The channel-adjusted model also showed significantly faster responses for visual than auditory probes.

\begin{table}[H]
\centering
\footnotesize
\renewcommand{\arraystretch}{0.95}
\caption{Mixed-effects models for Video 3 response times (log-transformed RT).}
\begin{tabular}{llrrrr}
\toprule
Model & Term & Coef. & SE & $z$ & $p$ \\
\midrule
\multirow{3}{*}{Total policy effect}
& Intercept & 6.489 & 0.040 & 161.45 & $<.001$ \\
& Model vs.\ Inverse & -0.178 & 0.059 & -3.02 & .003 \\
& Random vs.\ Inverse & -0.146 & 0.059 & -2.48 & .013 \\
\midrule
\multirow{4}{*}{Channel-adjusted}
& Intercept & 6.528 & 0.040 & 165.16 & $<.001$ \\
& Model vs.\ Inverse & -0.132 & 0.058 & -2.29 & .022 \\
& Random vs.\ Inverse & -0.114 & 0.058 & -1.98 & .048 \\
& Visual vs.\ Audio & -0.137 & 0.015 & -9.15 & $<.001$ \\
\bottomrule
\end{tabular}
\Description{Mixed-effects models for Video 3 show that routing policy significantly affects log-transformed response time. Relative to Inverse, both Model and Random produce faster responses. In the channel-adjusted model, visual probes are also significantly faster than auditory probes, while the faster response pattern for Model and Random remains significant.}
\label{tab:video3_mixedlm}
\end{table}

\subsection{Quality of the signal}

Our website at \url{https://dinithipurna.github.io/HeadRoom/} shows an example availability trace from one of our live Aria recordings. For readability, the visual and auditory availability signals are smoothed, but the figure still illustrates that the traces vary over time in intuitively expected ways as scene conditions change. Our companion website hosts a real-time video playback for this example. Table~\ref{tab:feature_correlations} shows the correlation results between our predicted channel availability and low-level features.

\begin{table}[H]
\centering
\footnotesize
\renewcommand{\arraystretch}{0.95}
\caption{Pearson correlations between handcrafted low-level features and \textsc{HeadRoom} availability estimates. The largest association was between motion energy and visual availability, while all other correlations were small.}
\begin{tabular}{lr}
\toprule
Feature pair & Pearson $r$ \\
\midrule
motion\_energy vs.\ visual availability   & -0.309 \\
luminance vs.\ visual availability        & -0.195 \\
scene\_cut vs.\ visual availability       & -0.178 \\
edge\_density vs.\ visual availability    & \phantom{-}0.111 \\
spectral\_flux vs.\ auditory availability & -0.071 \\
rms\_energy vs.\ auditory availability    & -0.065 \\
speech\_activity vs.\ auditory availability & -0.063 \\
rms\_energy vs.\ visual availability      & \phantom{-}0.058 \\
spectral\_flux vs.\ visual availability   & \phantom{-}0.056 \\
scene\_cut vs.\ auditory availability     & \phantom{-}0.049 \\
speech\_activity vs.\ visual availability & \phantom{-}0.032 \\
motion\_energy vs.\ auditory availability & \phantom{-}0.028 \\
edge\_density vs.\ auditory availability  & -0.012 \\
luminance vs.\ auditory availability      & -0.010 \\
\bottomrule
\end{tabular}
\Description{Correlations between low-level features and predicted availability are generally small. The largest association is a modest negative correlation between motion energy and visual availability (r = -0.309). Other correlations, including luminance, scene cuts, edge density, spectral flux, RMS energy, and speech activity, are weak, suggesting the availability estimates are not dominated by any single low-level feature.}
\label{tab:feature_correlations}
\end{table}









\end{document}